\documentclass[prd]{revtex4}
\usepackage{amsfonts}
\usepackage{amsmath}
\usepackage{amssymb}
\usepackage{graphicx}

\providecommand{\U}[1]{\protect\rule{.1in}{.1in}}
\newcommand{\ba}{\begin{eqnarray}}
\newcommand{\ea}{\end{eqnarray}}
\def\beq{\begin{equation}}
\def\eeq{\end{equation}}

\begin{document}

\begin{titlepage}
\begin{flushright}
ACT-01/09 \\
CERN-TH-PH/2008-005\\
MIFP-09/03 \\
January 2009
\end{flushright}

\vspace*{1cm}

\begin{centering}
{\bf {\large Probing a Possible Vacuum Refractive Index with $\gamma$-Ray Telescopes$^{*}$}}
\vspace{1cm}

{\bf John Ellis}$^a$, {\bf N. E. Mavromatos}$^b$,
{\bf D.V. Nanopoulos}$^{c,d,e}$

\end{centering}

\vspace{1cm}

\begin{centering}

{\bf Abstract}

\end{centering}

\vspace{0.5cm}
We have used a stringy model of quantum space-time foam to
suggest that the vacuum may exhibit a non-trivial refractive index
depending linearly on $\gamma$-ray energy: $\eta -1 \sim E_\gamma/M_{QG1}$,
where $M_{QG}$ is some mass scale typical of quantum gravity that may be $\sim 10^{18}$~GeV:
see~\cite{emnnewuncert} and references therein. The MAGIC, HESS and Fermi $\gamma$-ray telescopes
have recently probed the possible existence of such an energy-dependent vacuum refractive index.
All find indications of time-lags for higher-energy photons, but cannot exclude the
possibility that they are due to intrinsic delays at the sources.
However, the MAGIC and HESS observations of time-lags in emissions from AGNs Mkn 501 and
PKS 2155-304 are compatible with each other and a
refractive index depending linearly on the $\gamma$-ray energy, with $M_{QG1} \sim 10^{18}$~GeV.
We combine their results to estimate the time-lag $\Delta t$ to be expected for the
highest-energy photon from GRB 080916c measured by the Fermi telescope, which has an energy
$\sim 13.2$~GeV, assuming the redshift $z = 4.2 \pm 0.3$
measured by GROND. In the case of a refractive index depending linearly on the
$\gamma$-ray energy we predict
$\Delta t = 25 \pm 11$~s. This is compatible with the time-lag $\Delta t \le 16.5$~s reported by the
Fermi Collaboration, whereas the time-lag would be negligible in the case of a
refractive index depending quadratically on the $\gamma$-ray energy.
We suggest a strategy for future observations that could distinguish between a quantum-gravitational
effect and other interpretations of the time-lags observed by the MAGIC, HESS and Fermi
$\gamma$-ray telescopes.\\
~\\
~\\
{\bf $^*${\it Addendum} to `Derivation of a Vacuum Refractive Index in a Stringy Space-Time Foam
  Model', Phys.\ Lett.\  B 665, 412 (2008).}

\vspace{1cm}

\vspace{2cm}
\begin{flushleft}
$^a$ Theory Division, Physics Department, CERN, CH-1211, Geneva 23, Switzerland. \\
$^b$ Theoretical Physics, Department of Physics, King's College
London, Strand, London WC2R 2LS, U.K. \\
$^c$ Department of Physics,
Texas A \& M University, College Station,
TX~77843-4242, U.S.A. \\
$^d$ Astroparticle Physics Group, Houston
Advanced Research Center (HARC), Mitchell Campus, Woodlands, \\
TX 77381, U.S.A. \\
$^e$ Academy of Athens,
Chair of Theoretical Physics,
Division of Natural Sciences, 28~Panepistimiou Avenue,\\
Athens 10679, Greece.
\end{flushleft}
\end{titlepage}

The advent of the new generation of ground-based {\v C}erenkov $\gamma$-ray telescopes,
represented by MAGIC~\cite{MAGIC} and HESS~\cite{hess2155}, together with the launch of
the Fermi (n{\' e}e GLAST) $\gamma$-ray telescope~\cite{glast}, have inaugurated a new
era in $\gamma$-ray astronomy. It has been suggested that,
in addition to high-energy astrophysics, such
instruments may be used to probe fundamental physics. Specifically, data from
$\gamma$-ray telescopes may be used to probe the existence of a possible
energy-dependent vacuum refractive index~\cite{AEMNS}, such as might be induced by
quantum-gravitational effects in space-time foam: see~\cite{emnnewuncert} and references therein.

Since the publication of~\cite{emnnewuncert}, each of the
MAGIC, HESS and Fermi Collaborations has
reported time-lags in the arrival times of high-ernergy photons, as compared with
photons of lower energies. Following an earlier publication by the MAGIC Collaboration~\cite{MAGIC},
the present authors, A.~S.~Sakharov and E.~K.~G.~Sarkisyan joined the MAGIC
Collaboration in a quantitative analysis~\cite{MAGIC2} of the arrival times of individual photons from
the AGN Mkn 501, which has a redshift $z = 0.034$, finding an indication of an
energy-dependent time-lag
$\Delta t/E_\gamma = 0.030 \pm 0.012$~s/GeV. More recently, the HESS Collaboration
has published a cross-correlation analysis of the AGN PKS 2155-304~\cite{hessnew},
which has a redshift $z = 0.116$, finding a time-lag
$\Delta t = 28 \pm 25$~s between the light curves for $E_\gamma$ in the energy
ranges $(200, 800)$ and
$> 800$~GeV. Finally, the Fermi Collaboration has made a preliminary report of a
$4.5$-second time delay between the onsets of high- ($> 100$~MeV) and low-energy
($> 100$~KeV) emissions,
and a time-lag $\Delta t = 16.5$~s for a photon with energy $13.2^{+0.70}_{-1.54}$~GeV
from the GRB 080916c~\cite{grbglast}.

These observations provide a basis for discussing
possible interpretations of time-lags in the arrival times of energetic photons, and
provide pointers for possible future analyses that we discuss in this {\it addendum} to~\cite{emnnewuncert}.

The most conservative interpretations of the MAGIC, HESS and Fermi time-lags
are that they are due to emission mechanisms at the sources.
The mechanisms leading to $\gamma$-ray
emissions from AGNs and GRBs are surely different. However, it is possible that there may be
some common systematic effect leading to higher-energy photons being emitted later. For instance,
it has been suggested~\cite{grbglast} that it may be easier and quicker for sources to accelerate
electrons, which are relatively light and therefore may produce
the early, lower-energy part of these bursts, whilst sources may take longer to accelerate
the heavier protons, which would then contribute later to the higher-energy component.
In addition to any such common systematic time-lag, there may be different intrinsic time-lag
effects present in AGNs and GRBs, or even between different classes of GRBs that
have different underlying mechanisms (e.g., long bursts vs short bursts).
There may also be intrinsic fluctuations in the relative emission times of higher- and lower-energy
components between different sources of the same class, or even between different features
in the time structure of a single source. These different possibilities limit the sensitivities of
probes for quantum-gravity effects, and it is clear that they must all be taken into account and
controlled before claiming a robust lower limit on any observable effect of
space-time foam, and {\it a fortiori} before claiming the existence of any such effect.

With these points in mind, an analysis of a significant sample of long-burst
GRBs was performed in~\cite{robust}, extending earlier analyses~\cite{mitsou}. An attempt was made to distinguish intrinsic
time-lags at the sources (which were assumed to be independent of redshift) from
any time-lag induced by a possible quantum-gravitational effect during propagation,
which would be redshift-dependent. In the case of a linear effect on propagation
characterized by an energy scale $M_{QG1}$,
the relative time-lag of a higher-energy photon is
\begin{equation}
\Delta t_{\rm no~expansion} \sim \frac{ \Delta E}{M_{\rm QG1}} \frac{L}{c}
\label{naive}
\end{equation}
at small redshifts, where $\Delta E$ is the difference in photon energies, $c$ is the speed of light in vacuo, $L = H_0 z c$ is the distance of the source from Earth, and
${H}_0 \sim 2.5 \times 10^{-18} ~{\rm s}^{-1}$ is the current Hubble expansion rate. For
sources at a cosmological distance, one must take into account the expansion history
of the Universe. We assume that this is described adequately by the standard
$\Lambda$CDM model of cosmology, so that~\cite{mitsou,robust,JP}
 \begin{equation}
(\Delta t)_{\rm obs} =  \frac{ \Delta E}{M_{\rm QG1}} {\rm H}_0^{-1}\int_0^z \frac{(1 + z)dz}{\sqrt{\Omega_\Lambda + \Omega_m (1 + z)^3}} ,
\label{redshift}
\end{equation}
and we assume for concreteness $\Omega_\Lambda \sim 0.73$ and $\Omega_m \sim 0.27$.

The robust analysis in~\cite{robust} of a sample of GRBs extending to large redshifts found
indications of a red-shift-independent intrinsic time-lag as well as fluctuations in the time-lags
between different structures. After controlling for these effects, and performing a linear
regression analysis in terms of the appropriate function of the redshift:
\begin{equation}
K(z) \equiv \int_0^z \frac{(1 + z)dz}{\sqrt{\Omega_\Lambda + \Omega_m (1 + z)^3}} ,
\label{K}
\end{equation}
which reduces to $z$ in the small-redshift limit,
no significant evidence was found for a redshift-dependent propagation effect, and a lower bound
$M_{QG1} > 1.4 \times 10^{16}$~GeV was found at the 95\% confidence level. It should be
noted that this analysis used photons with energies $< 320$~KeV, much below those
observed by the $\gamma$-ray telescopes discussed here, and that the intrinsic time-lags
and their fluctuations were at the level of 0.1~s or less. 

It was not possible to control for intrinsic effects in the analysis of
MAGIC data in~\cite{MAGIC2}, which was therefore not as robust as the previous 
GRB analysis~\cite{robust},
as was made clear in quoting the lower limit $M_{QG1} > 2.1 \times 10^{17}$~GeV
at the 95\% confidence level~\cite{MAGIC2}. The same remark applies to the lower limit
$M_{QG1} > 7.2 \times 10^{17}$~GeV at the 95\% confidence level found subsequently
in a cross-correlation analysis of HESS data~\cite{hessnew}.

The available sample of AGNs is still not large enough for a robust
regression analysis. However, one can at least check for consistency between the
available MAGIC and HESS results, and gauge the magnitude of possible intrinsic
fluctuations in the AGN time-lags. Comparing the time-lag measured by MAGIC for
Mkn 501at $z = 0.034$: $\Delta t/E_\gamma = 0.030 \pm 0.012$~s/GeV, with that
measured for PKS 2155-304 at $z = 0.116$: $\Delta t/E_\gamma = 0.030 \pm 0.027$~s/GeV,
we see that they are compatible with a common, energy-dependent {\it intrinsic} time-lag
at the source. On the other hand, neglecting possible source effects, the one-$\sigma$ range for the MAGIC
time-lag corresponds to $M_{QG1} = (0.48^{+0.32}_{-0.14}) \times 10^{18}$~GeV, which
is compatible with the HESS 95\% C.L. lower limit of $0.72 \times 10^{18}$~GeV~\cite{Wagner}.
The MAGIC and HESS data are compatible with a universal redshift- and energy-dependent
{\it propagation} effect:
\begin{equation}
\Delta t/E_\gamma = (0.43 \pm 0.19) \times K(z) {\rm s/GeV} ,
\label{bestfit}
\end{equation}
corresponding to $M_{QG1} = (0.98^{+0.77}_{-0.30}) \times 10^{18}$~GeV~\cite{Whipple}.
Discriminating between these respective
{\it conservative} and {\it audacious} interpretations will require considerably more data from sources
at different redshifts emitting in different
energy ranges. It seems unlikely that the relatively rare
and unpredictable sharp energetic flares produced
only occasionally by AGNs, which have a relatively restricted redshift range and hence a
small lever arm, will soon be able to provide the desired discrimination.

On the other hand, GRBs are observed at a relatively high rate, about one a day, and
generally have considerably larger redshifts. The AGN data cannot be used to estimate
the possible magnitudes of intrinsic time-lags in GRB emissions, since the sources are
very different. Nor can the previous GRB data mentioned earlier be used to estimate the
likely GRB time-lags above the KeV range. However, the above-mentioned best fit to
the possible redshift- and energy-dependent propagation effect, (\ref{bestfit}), does
provide an estimate of the sensitivity required to probe such an effect in GRB data in the
GeV range.

The advent of the Fermi (n{\' e}e GLAST) telescope with its large acceptance offers the
possibility of achieving the required sensitivity. Indeed, the Fermi Collaboration has
already made a preliminary report of GeV-range $\gamma$ rays from the GRB 080916c.
As already mentioned, there is a 4.5-second time-lag between the onsets
of high- ($> 100$~MeV) and low-energy ($< 100$~KeV)
emissions. Moreover, the highest-energy photon GRB 080916c measured by the Fermi
$\gamma$-ray telescope had an energy $E = 13.2^{+0.70}_{-1.54}$~GeV, and was
detected $\Delta t = 16.5$~s after the start of the burst.
Spectroscopic information has been used by the GROND Collaboration~\cite{GROND} to estimate
the redshift of GRB 080916c as $z = 4.2 \pm 0.3$~\cite{grbglast}. Assuming
this value of the redshift, the best fit (\ref{bestfit}) would correspond to
a time-lag
\begin{equation}
\Delta t = 25 \pm 11~{\rm s}
\end{equation}
for a 13~GeV photon from GRB 080916c.
The consistency between this Fermi measurement and the best-fit estimate (\ref{bestfit})
is \emph{striking}, but one should keep in mind that all or part of the 16.5~s time-lag
of this highest-energy photon could be due to intrinsic effects. Indeed, the 4.5-second time-lag
observed for $\sim 100$~MeV photons could not be explained by a propagation
effect that depends linearly on the energy. Because of ignorance of the source
mechanism, the preliminary analysis of these data by the Fermi
Collaboration~\cite{grbglast} quoted a lower bound
$M_{QG1} > (1.50 \pm 0.20) \times 10^{18}~{\rm GeV}$, which is consistent with the MAGIC
and HESS results stated previously. It is clear that the Fermi
telescope has already demonstrated the sensitivity to probe a possible linearly
energy-dependent propagation effect at the level reached by the available AGN data,
and it is appropriate and possibly helpful to consider how such an effect could be probed
in the future.

If the apparent consistency between the available AGN and GRB data would persist,
it would provide much more convincing evidence for a possible linearly energy-dependent
propagation effect than could either AGN or the GRB data alone, since the sources are so
different in nature. However, `extraordinary claims require extraordinary evidence',
so for the moment we can applaud the efforts of the MAGIC, HESS and Fermi
Collaborations to date, wish them good fortune in the future, and stress the advantages
of a combined analysis of AGNs and GRBs.

It is interesting to repeat the above analysis for the
case of a possible quadratically energy-dependent effect. We recall that the MAGIC
analysis found $\Delta t/E_\gamma^2 = (3.7 \pm 2.6) \time 10^{-6}$~s/GeV$^2$,
whereas the HESS result corresponds to
$\Delta t/E_\gamma^2 = (3.3 \pm 2.9) \time 10^{-5}$~s/GeV$^2$. (These numbers
were estimated using the HESS statement that the average energy difference between
photons in the high- and low-energy bins used in their analysis was 0.92~TeV. A more
precise number could be obtained by analyzing the arrival times and energies of individual photons.)
These results are compatible, but neither has any indication of a non-zero result. The MAGIC result,
which is considerably more sensitive, would correspond to
$\Delta t/E_\gamma^2 = (1.4 \pm 1.0) \times 10^{-3}$~s/GeV$^2$ for energetic
$\gamma$-rays from GRB 080916c at $z = 4.2$. This would imply a time delay of $0.24 \pm 0.16$~s
for the most energetic photon measured by the Fermi telescope, two orders of magnitude less
than the time-lag measured. Of course, the observed time-lag could be due solely to intrinsic effects
at the source. Nevertheless, it is intriguing that a possible linear energy-dependent effect
{\it could} explain simultaneously all three sets of data, whereas a quadratic effect
{\it could not}.

This point emphasizes the value of a combined analysis of data from sources at
different redshifts and in different energy bands: in principle, they could not only distinguish
between source and propagation effects, but also between different energy dependences.

We conclude with some theoretical observations.

Many theories of quantum gravity predict non-trivial vacuum refractive indices, varying linearly with the energy scale of photons and with the distance of the source: see~\cite{aemn,mitsou,gambini,mestres,dsr,smolin,horizons,ems,myers,emnw,emnnewuncert}.
However, there are several stringent restrictions coming from other independent tests of Lorentz symmetry that should be taken into account before any model could be accepted as a
possible explanation for any observed photon delays.

Linear energy-dependent effects (with a quantum-gravity energy scale of the order of the Planck
mass scale or even larger) on the propagation of charged probes, such as electrons, are
already excluded by synchrotron radiation measurements~\cite{crab,crab2,ems}, and similar effects
for photons in models that entail birefringence are excluded by GRB afterglow
measurements~\cite{uv,grb}. This is a significant restriction, since most models of quantum
gravity that are based on a local effective Lagrangian do exhibit birefringence
(see e.g.~\cite{gambini,mestres,myers}). Moreover, most of these models are characterised by the absence of GZK thresholds for the extinction by microwave background photons
of ultra-high-energy photons with energies higher
than $10^{19}$~eV, due to electron-positron pair production.
The non-observation of such ultra-high-energy photons imposes severe restrictions on the
quantum-gravity linear suppression scale to values more than seven orders of magnitude higher
than the Planck scale~\cite{sigl}, which may be evaded only if there are energy fluctuations during particle interactions~\cite{emngzk}.
Another feature of many models is photon decay, which can become possible if modified dispersion relations for photons are present~\cite{sigl}.

It is clear from this summary that any model
of refraction in space-time foam that exhibits effects at the level of the MAGIC~\cite{MAGIC2},
HESS~\cite{hessnew} and Fermi sensitivity~\cite{grbglast} would be characterised by the
following \emph{specific} properties:

{(i)} photons are \emph{stable} (i.e. do \emph{not} decay) but should exhibit a
modified \emph{subluminal} dispersion relation with Lorentz-violating corrections that grow
linearly with $E/M_{QG1}$, where $M_{QG1}$ is close to the Planck scale,

{(ii)} the medium should not refract electrons, so as to avoid the synchrotron-radiation
constraints~\cite{crab,ems},

{(iii)} the coupling of the photons to the medium must be independent of photon polarization, so as
to avoid birefringence, thus avoiding the stringent pertinent constraints~\cite{uv,grb,crab2}, and,
moreover,

{(iv)} the formalism of a local effective field theory lagrangian
in an effectively flat space-time breaks down, including higher-derivative local
interaction terms to produce a refractive index~\cite{myers}, which
would be signalled by quantum fluctuations in the total energy in particle interactions,
due to the presence of a quantum-gravitational `environment'~\cite{emngzk}.

A model with all these properties has been suggested by us~\cite{horizons,emnw,ems}
within the framework of string/brane theory,
based on a stringy analogue of the interaction of a photon with internal degrees of freedom in a conventional medium. We modelled the space-time foam as a gas of point-like D-brane defects
(D-particles) in the bulk space-time of a higher-dimensional cosmology where
the observable Universe is a D3-brane.
Within this class of D-foam models, we have recently re-derived~\cite{emnnewuncert} a
refractive index for photon propagation in vacuo using
a detailed modelling of the interaction of an open string, representing a
photon, with a D-particle. During this interaction, an intermediate open string state is created,
which stretches between the D-particle and the D3-brane. The D-particle is excited in the process,
and subsequently decays re-emitting the photon, with a causal time delay that increases linearly
with the photon energy. The whole process is consistent with the stringy uncertainty principles,
as shown in detail in~\cite{emnnewuncert}.
An important feature of the model is that only electrically-neutral excitations can be captured by the
D-particles, whereas the D-particle foam looks transparent to charged particles such as electrons.
This is because the capture process entails~\cite{emnnewuncert} a splitting of the open-string
state representing matter excitations. Charged excitations are characterized by an electric flux
flowing across the string, and when the latter is cut in two pieces as a result of its
capture by the D-particle defect, the flux should go somewhere because charge is conserved.
For this reason, our stringy model avoids the stringent constraints coming from synchrotron
radiation~\cite{crab,crab2,ems}.
Moreover, the independence of the time delay from the photon polarization implies the absence of birefringence effects, thus avoiding the very strong constraints on such
Lorentz-violating effects inferred from GRB afterglows~\cite{grb,uv}. The model also
predicts energy fluctuations during particle interactions~\cite{emngzk}, thereby
evading the GZK constraint~\cite{sigl}.

The interactions of photons with the D-particle foam generate~\cite{emnnewuncert} an effective
subluminal refractive index $\eta(E)$ for light propagating in this space-time, since
light being slowed down by the medium effects. In fact, our model
of D-particle foam has many analogies with the simple harmonic oscillator model
discussed in~\cite{feynman} as a description of the refraction of light in conventional media.
The r\^ole of atomic electrons in the conventional model, represented by harmonic oscillators,
is played by the D-particle defects. However, being stringy in nature, they are characterised by
an infinity of oscillation modes. The role of the restoring force that keeps atomic electrons in
position during the scattering of light is played~\cite{emnnewuncert} by the intermediate-state
flux-carrying strings, that are stretched between the D-particle and the D3-brane during the
capture of the photon (represented as an open string with each end attached to the D3-brane)
by the D-particle.

It is not clear to us whether other candidates of space-time foam, in particular~\cite{dsr,smolin}
could satisfy simultaneously all the requirements (i)-(iii) mentioned above. For instance, although
there are claims~\cite{dsr,smolin} that there is no birefringence,
it is not clear whether these candidates are derived from local effective lagrangians and how they
could avoid the stringent constraints~\cite{sigl} imposed by the absence of ultrahigh-energy
photons and photon decays~\cite{sigl}. On the other hand, as a result of the presence of
D-particles in our model of space-time foam, and their recoils,
the local effective lagrangian description breaks down, and thus the stringent constraints
of~\cite{sigl} are evaded~\cite{emngzk}.

We close by repeating that, with measurements of only a few AGN flares
from MAGIC and HESS, it is
not possible to disentangle with any certainty intrinsic source effects from propagation effects.
For this one would need statistically significant populations of AGN data.
Unfortunately the occurrence of fast flares in AGNs is currently unpredictable,
and since no correlation has yet been established with observations in other
energy bands that could be used as a trigger signal, only serendipitous
detections are currently possible. However, the encouraging news from the GRB front,
in particular from the Fermi $\gamma$-ray telescope,
leads to the hope that there will soon be many more
observations of energetic photons from GRBs like 080916c, which could play an
increasingly important role in future quantum gravity tests. In particular, the different
redshifts of GRB data could help disentangle source and propagation effects, and
the different energy ranges of the GRB and AGN data could help distinguish between
different possible energy dependences.

\section*{Acknowledgements}

The work of J.E. and N.E.M. is partially supported by the European Union
through the Marie Curie Research and Training Network \emph{UniverseNet}
(MRTN-2006-035863), and that of D.V.N. by DOE grant DE-FG02-95ER40917.

\end{document}